\documentclass[aps,prl,twocolumn,superscriptaddress]{revtex4}

\usepackage{amsmath,bm}
\usepackage{graphicx}
\usepackage{upgreek}

\usepackage[retainorgcmds]{IEEEtrantools}

\renewcommand{\Re}{\operatorname{Re}}
\renewcommand{\Im}{\operatorname{Im}}

\newcommand{\Syout}{S_y^{\mathrm{out}}}
\newcommand{\Syin}{S_y^{\mathrm{in}}}
\newcommand{\var}{\mathrm{var}}

\newcommand{\LR}[1]{\left[  #1 \right]}
\newcommand{\Ttot}{T_{\mathrm{tot}}}
\newcommand{\Tdark}{T_2^{\mathrm{dark}}}

\newcommand{\fTrtHz}{\mathrm{fT}/\sqrt{\mathrm{Hz}}}

\newcommand{\lr}[1]{\left(  #1 \right)}
\newcommand{\abs}[1]{\left|  #1 \right|}
\newcommand{\rtHz}{\sqrt{\mathrm{Hz}}}

\newcommand{\bracket}[1]{\left\langle #1 \right\rangle}
\newcommand{\braket}[1]{\left\langle #1 \right\rangle}
\newcommand{\Bnerve}{B_{\mathrm{nerve}}}

\newcommand{\BPN}{B_{\mathrm{PN}}}

\newcommand{\Navg}{N_{\mathrm{avg}}}

\newcommand{\vecB}{\mathbf{B}}
\newcommand{\vecJ}{\mathbf{J}}
\newcommand{\HB}{\mathcal{H}_B}
\newcommand{\gyromag}{\gamma}
\newcommand{\g}{\Gamma}
\newcommand{\gdark}{\Gamma_{\mathrm{dark}}}
\newcommand{\um}{\mu \mathrm{m}}

\begin{document}
 
\title{Non-invasive detection of animal nerve impulses with an atomic magnetometer operating near quantum limited sensitivity.}

\author{Kasper Jensen}
\affiliation{Niels Bohr Institute, University of Copenhagen, Blegdamsvej 17, 2100 Copenhagen, Denmark}
\author{Rima Budvytyte}
\affiliation{Niels Bohr Institute, University of Copenhagen, Blegdamsvej 17, 2100 Copenhagen, Denmark}
\author{Rodrigo A. Thomas}
\affiliation{Niels Bohr Institute, University of Copenhagen, Blegdamsvej 17, 2100 Copenhagen, Denmark}
\author{Tian Wang}
\affiliation{Niels Bohr Institute, University of Copenhagen, Blegdamsvej 17, 2100 Copenhagen, Denmark}
\author{Annette Fuchs}
\affiliation{Department of Biomedical Sciences, Faculty of Health and Medical Sciences, University of Copenhagen, Blegdamsvej 3, 2200 Copenhagen N, Denmark}
\author{Mikhail V. Balabas}
\affiliation{Niels Bohr Institute, University of Copenhagen, Blegdamsvej 17, 2100 Copenhagen, Denmark}
\affiliation{Department of Physics, St Petersburg State University, Universitetskii pr. 28, 198504 Staryi Peterhof, Russia}
\author{Georgios Vasilakis}
\affiliation{Niels Bohr Institute, University of Copenhagen, Blegdamsvej 17, 2100 Copenhagen, Denmark}
\author{Lars Mosgaard}
\affiliation{Niels Bohr Institute, University of Copenhagen, Blegdamsvej 17, 2100 Copenhagen, Denmark}
\author{Thomas Heimburg} 
\affiliation{Niels Bohr Institute, University of Copenhagen, Blegdamsvej 17, 2100 Copenhagen, Denmark}
\author{S{\o}ren-Peter Olesen}
\affiliation{Department of Biomedical Sciences, Faculty of Health and Medical Sciences, University of Copenhagen, Blegdamsvej 3, 2200 Copenhagen N, Denmark}
\author{Eugene S. Polzik}
\affiliation{Niels Bohr Institute, University of Copenhagen, Blegdamsvej 17, 2100 Copenhagen, Denmark}

\maketitle

%%%%%%%%%%%%%%%%%%%%%%%%%%%%%%%%%%%%%%%%%%%%%%%%%%%%%%%

\textbf{Magnetic fields generated by human and animal organs, such as the heart, brain and nervous system carry information useful for biological and medical purposes. These magnetic fields are most commonly detected using cryogenically-cooled superconducting magnetometers. Here we present the first detection of action potentials from an animal nerve using an optical atomic magnetometer. Using an optimal design we are able to achieve the sensitivity dominated by the quantum shot noise of light and quantum projection noise of atomic spins. Such sensitivity allows us to measure the nerve impulse with a miniature room-temperature sensor which is a critical advantage for biomedical applications. Positioning the sensor at a distance of a few millimeters from the nerve, corresponding to the distance between the skin and nerves in biological studies, we detect the magnetic field generated by an action potential of a frog sciatic nerve. From the magnetic field measurements we determine the activity of the nerve and the temporal shape of the nerve impulse. This work opens new ways towards implementing optical magnetometers as practical devices for medical diagnostics.
}

The magnetic field generated around a signaling nerve fiber is of key interest both from a basic scientific and a clinical point of view. 
The transmembrane potentials have been extensively measured with electrophysiological techniques.
Magnetic field measurements are insensitive to the transmembrane currents  as the fields from the opposite currents in and out of the membrane cancel.
Instead, magnetic field measurements allow for  a true measurement of the axon's axial net current, which is the depolarizing wavefront driving the action potential. Magnetic field recordings also allow for non-invasive measurements of the conduction velocity of peripheral nerves which is necessary for diagnostics of multiple sclerosis, myotonia and intoxication in patients.

The magnetic field of a nerve impulse was first measured by Wikswo et al \cite{Wikswo1980} using a combination of  a superconducting SQUID magnetometer and a toroidal pick-up coil through which the nerve had to be pulled. This method is not compatible with in vivo diagnostics and yields the magnetic field values which are much higher than that in an animal because the return currents in the surrounding tissue are not measured. 
Here we are able to detect the nerve impulse with the sensor placed beside the nerve, several millimeters away, the setting compatible with in vivo studies.

Sensitivity of atomic magnetometers \cite{Budker07} improves with the number of atoms sensing the field, which for vapor magnetometers is defined by volume and temperature. For example, femtoTesla sensitivity has been achieved with magnetometers operating at a temperature of several hundred of $^\circ$C in the so-called SERF regime \cite{Kominis2003nature} used also  for medical applications \cite{Alem2015PhysMedBio,Sander2012biomedexpress,Alem2014neuro}.
Similarly high sensitivity has been achieved  at room temperature using much fewer atoms by means of quantum state engineering \cite{Wasilewski2010prl} leading to operation beyond standard quantum limits of sensitivity.  Room temperature operation allows to place the sensor in contact with the skin or potentially inside the human body. The close proximity of the sensor to the source of magnetic field is a big advantage as the magnetic field rapidly decreases with the distance from the source. Room-temperature cesium magnetometer has been used for medical applications \cite{Bison2009apl}, however, it operated far above quantum limits of sensitivity. 

\begin{figure*}[ht]
\centering
\includegraphics[width=1\textwidth]{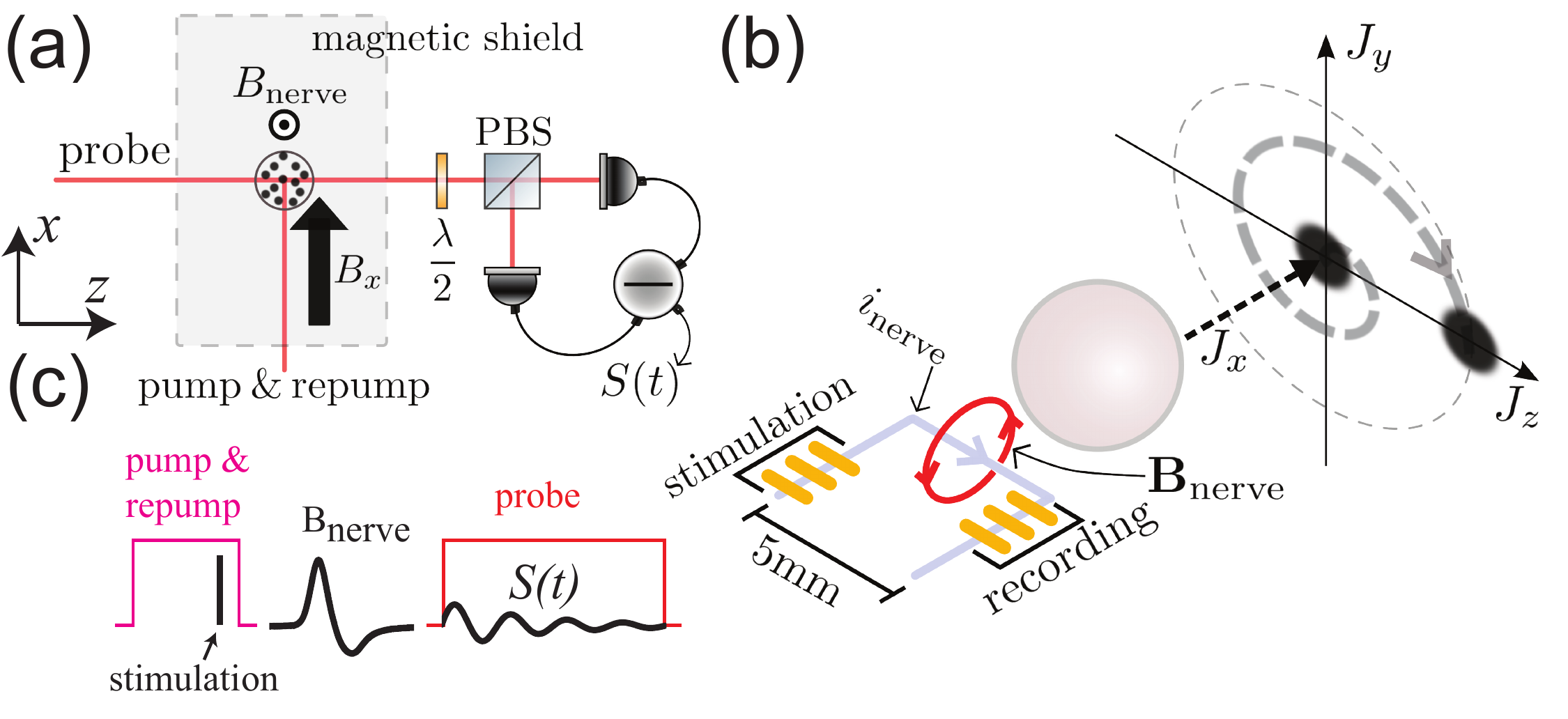}
\caption{(a) Schematic of the experimental setup. Probe light propagates along the $z$ axis. Half-wave plate $\lambda/2$, polarizing beam splitter (PBS) and differential photodetection are components of polarization detector. (b) The magnetometer principle. The amplitude of the collective atomic spin precession in $z,y$ plane is proportional to $B_{\rm{nerve}}$. Spin projection $J_z$ is measured by probe light with the sensitivity limited by the quantum projection spin noise (fuzzy circle). (c) The measurement sequence for the pulsed magnetometer mode.
}
\label{fig:setup}
\end{figure*}

Here we use the approach of Ref.~\cite{Wasilewski2010prl} for nerve impulse measurements. The sensitive element of the magnetometer is cesium atomic vapour. Cesium has a high vapor pressure such that high sensitivity can be reached at room- or human body temperature. The magnetic moment (spin) of atoms $\vecJ=\lr{J_x,J_y,J_z}$ is prepared by optical pumping in the $x$-direction, along the direction of a bias field $B_x$ [see Fig.~\ref{fig:setup}(a)]. 
The magnetic field of the nerve $\Bnerve$ will create a transverse spin component $\vecJ_{\perp} = \lr{J_y,J_z}$ which afterwards will rotate in the $y$-$z$ plane at the Larmor frequency $\Omega=B_x/\gyromag$ [see Fig.~\ref{fig:setup}(b)], where $\gyromag=2.20 \cdot 10^{10}$~rad/$\lr{\mathrm{s}\cdot \mathrm{T}}$  is the cesium gyromagnetic ratio. 
The $J_z$ spin component  is detected optically by measuring the polarization rotation of  the probe light.  
The magnetic field from the nerve is detected in two modalities,
a continuous mode where the magnetic field as a function of time  $B(t)$  is detected, and a pulsed mode where the Fourier component $| B(\Omega) |$ is detected.
In the continuous mode the pump and probe light is continuously on. 
In the pulsed mode [see Fig.~\ref{fig:setup}(c)], a pulse of pump light is followed by the pulse of magnetic field, and finally the spins are detected with a pulse of probe light. 

Optical magnetometers are fundamentally limited by quantum spin-projection noise (PN) shown as the fuzzy circle in Fig.~\ref{fig:setup}(b). This limit has been reached for magnetic fields oscillating at hundreds of kHz in \cite{Wasilewski2010prl}.  The techniques allowing us to approach the PN limited sensitivity for nerve impulses whose frequency is much lower are described in Supplementary Material.
For continuous measurements, the PN limited magnetic field uncertainty $\Delta \BPN$ normalized by the total measurement time $\Ttot$ yields the sensitivity
$\Delta \BPN \sqrt{\Ttot} \sim 1/ \lr{\gyromag \sqrt{T_2 J_x /2}}$ 
in units of T/$\rtHz$. 
$T_2$ is the spin coherence time and $ J_x = 4 N_A$  is the total atomic spin for $N_A$ cesium atoms.  
At room temperature of $22^\circ$C, the cesium atomic density is $3.6 \times 10^{16}$~$\mathrm{m}^{-3}$   which is the highest of all elements appropriate for atomic magnetometry. 
The pulsed measurement has the PN-limited uncertainty of 
$\Delta \abs{B_{\mathrm{PN}}(\Omega)} = 1/\lr{\gamma \sqrt{2J_x}}$ if the magnetic  pulse duration  $\tau \ll T_2$ \cite{sup}. 

\begin{figure*}[ht]
\centering
\includegraphics[width=1\textwidth]{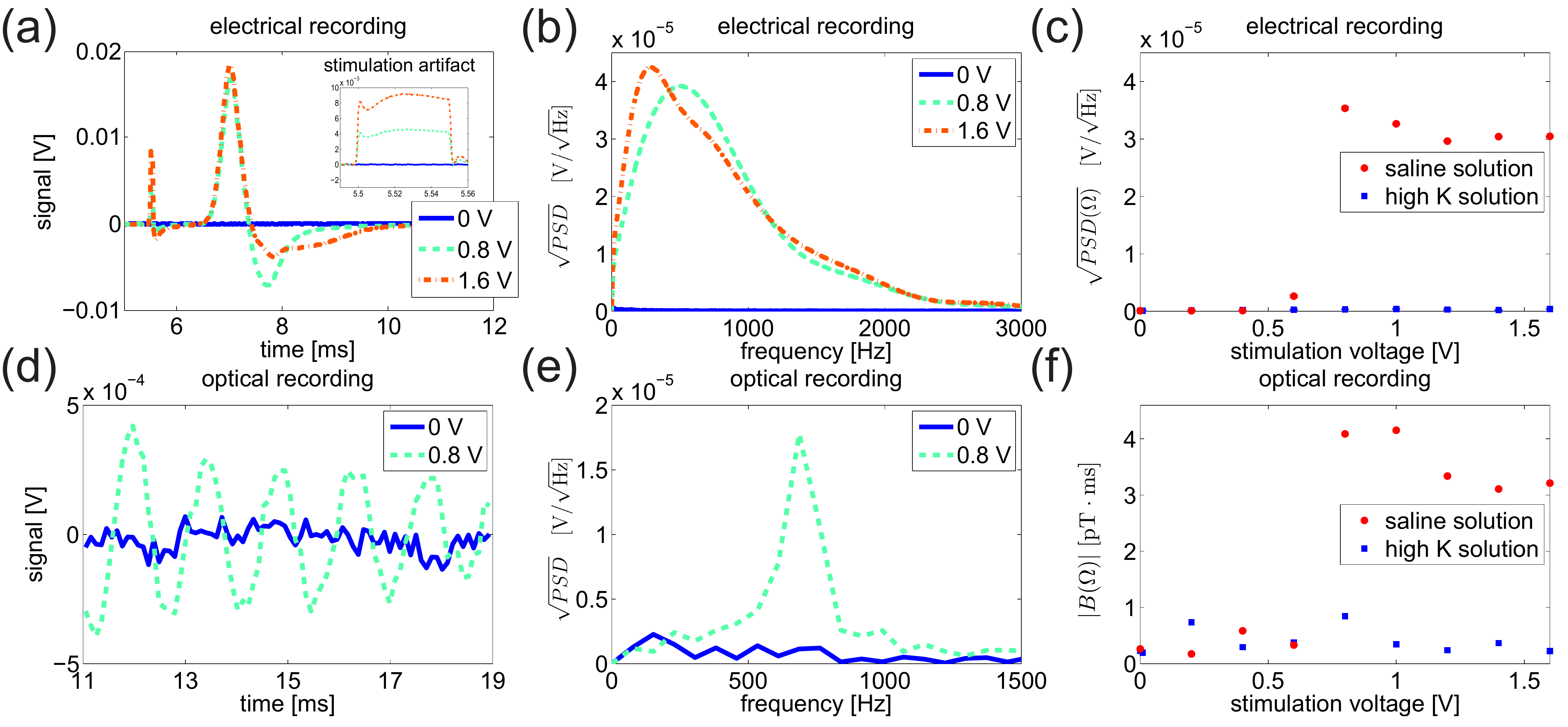}
\caption{ Electrical and optical measurements of the nerve impulse for different stimulation voltages. 
The optical measurements were done in the pulsed mode using 1000 averages.The figures show the signals in time, the square-root of the power spectral density $PSD$ and the 700~Hz frequency component. The plotted electrical signals are after 10 times amplification. The uncertainties on the data points in (c) are to small to be visible in the figure. The uncertainties on the points in (f) can be estimated as the standard deviation (0.20~pT$\cdot$ms) of the data points where the nerve did not fire. By dividing the nerve signal  
(4.1~pT$\cdot$ms)  by the standard deviation we find the signal to noise ratio, $SNR\approx 20$. }
\label{fig:pulsedelectricaloptical}
\end{figure*}

A long spin-coherence time $T_2$ is crucial for a high sensitivity. In this work we utilize a vapor cell with the inside surface coated with alkane  \cite{Balabas2010oe,Balabas2010prl}.  The coating protects atomic spin states from decoherence over many thousands of wall collisions and provides $T_2^{\mathrm{dark}}=15$~ms which is longer than a typical nerve impulse duration $\tau\approx2$~ms, as required for the ultimate sensitivity. The inner diameter of the  vapour cell is  5.3~mm and a wall thickness of 0.85~mm allows us to have atoms at an average distance of $4$ mm from the nerve axis which is close to a typical distance for many medical applications. 

A frog sciatic nerve contains a few nerve bundles each with several thousand axons inside (see Methods section). The nerve is placed inside a plastic chamber where it can be kept alive in a saline solution for more than 5 hours.  The nerve is electrically stimulated from one end with a pair of gold electrodes [see Fig.~\ref{fig:setup}(b)]. The stimulus triggers an action potential (a nerve impulse) propagating along the nerve. As a reference measurement we perform an electrical recording of the impulse with another pair of  electrodes.
Figure~\ref{fig:pulsedelectricaloptical}(a) shows the electrically recorded signals for different  stimulation voltages.  Figure~\ref{fig:pulsedelectricaloptical}(b) shows the frequency spectra of the nerve signals and  Fig.~\ref{fig:pulsedelectricaloptical}(c) shows the amplitude of the 700~Hz Fourier component. 
The nerve is stimulated at $t=5.5$~ms. The signature of the nerve signal  is its  non-linear behavior with stimulation voltage, with a firing threshold at around  0.7~V. For voltages above the threshold a nerve impulse is measured with the recording electrodes within the time interval $t=6.5-9.5$~ms.
We also observe a stimulation artifact at $t=5.5$~ms [see inset in Fig.~\ref{fig:pulsedelectricaloptical}(a)] which is proportional to the stimulation voltage. 

In parallel to the reference electrical recording the nerve signal is detected optically using the pulsed magnetometer mode.
The magnetometer is positioned near the middle part of the nerve separated only by a thin microscope cover slip.
As the nerve is bent in a U-shape, we mainly detect the field from the 5~mm section of the nerve closest to the magnetometer [see Fig.~\ref{fig:setup}(b)]. 
The axial ionic  current in this 5~mm section can be considered constant, as the action potential has a duration  $\approx 2$~ms, the velocity $\approx~40$~m/s, and therefore an extent $\approx 8$~cm $\gg5$~mm.
The circumferal magnetic field $\mathbf{B}_{\mathrm{nerve}}$ from the nerve is on average transverse to the initial spin direction $J_x$ and will therefore create a transverse spin component [see Fig.~\ref{fig:setup}(b)].
Figure~\ref{fig:pulsedelectricaloptical}(d) shows the magnetometer signal, Fig.~\ref{fig:pulsedelectricaloptical}(e) shows the spectrum and Fig.~\ref{fig:pulsedelectricaloptical}(f) shows the magnetic field Fourier component at the Larmor frequency of 700~Hz. A clear threshold for the nerve firing is observed confirming that the magnetometer is capable of detecting the nerve impulse.
From calibration measurements (see Methods section) we determine the Fourier component of the magnetic field from the nerve as
$\abs{\Bnerve\lr{\Omega}}= 4.1$~pT$\cdot$ms.
Note that the stimulation artifact which was observed in the electrical recording is not detected with the pulsed magnetometer as the stimulation occours during the optical pumping  [see Fig.~\ref{fig:setup}(c)] where the response to magnetic fields is strongly damped.

As a control experiment, we make the nerve inexcitable \cite{Krnjevic1954JPhysiology} by replacing the saline solution in the plastic chamber with a solution with high potassium concentration.
As expected, we clearly observe from both electrical [Fig.~\ref{fig:pulsedelectricaloptical}(c)] and optical [Fig.~\ref{fig:pulsedelectricaloptical}(f)] measurements that the nerve signal is blocked by this solution.
From Fig.~\ref{fig:pulsedelectricaloptical}(f) we infer that the nerve impulse can be detected optically with a signal to noise ratio $SNR\approx20$ using 1000 averages.
As the $SNR$ scales as $1/\sqrt{\Navg}$ we find the $SNR\approx 0.6$ for a single shot, i.e., we should be able to detect a nerve impulse in a single shot with just a minor improvement in $SNR$. 

\begin{figure*}[ht]
\centering
\includegraphics[width=1\textwidth]{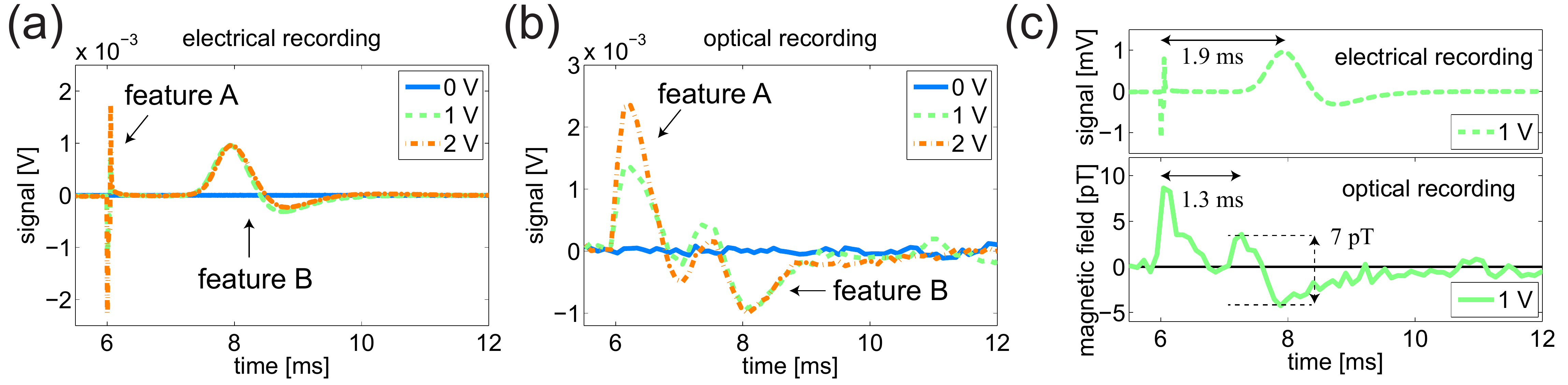}
\caption{Electrical and optical measurements of the nerve impulse for different stimulation voltages. The magnetometer was operated in the continuous mode and the signals were averaged 5000 times. (a) Electrical and (b) optical measurements. (c) Optical signal for 1 V stimulation and magnetic field calculated by deconvolution. 
 For these specific measurements the Larmor frequency was 410~Hz and the coherence time 0.44~ms.}
\label{fig:contmeas}
\end{figure*}
% Real numbers: Larmor frequency was 414 Hz and T2 was 0.44 ms.

The magnetometer can also be operated in the continuous mode which allows for determination of the temporal shape of the magnetic field generated by the nerve, $\Bnerve(t)$. The magnetometer response was optimized by matching its frequency response (a Lorentzian centered at the Larmor frequency  with a full width at half maximum $1/\lr{\pi T_2}$) with the spectrum of the nerve impulse (see Fig.~\ref{fig:pulsedelectricaloptical}(b)).
The bandwidth $1/\lr{\pi T_2}=720$~Hz was set by choosing suitable power levels for the pump, repump and probe lasers.
 Figures~\ref{fig:contmeas}(a) and \ref{fig:contmeas}(b) show the electrical and optical signals respectively as a function of time for different stimulation voltages. In both electrical and optical measurements we observe two features (A and B). Feature A is due to the stimulation as it starts at the time of stimulation and increases  linearly with the stimulation voltage. 
Feature B is due to the nerve signal, as it starts a few ms after the stimulation and only appears above the threshold for the nerve firing (at 1~V or greater). 
Figure~\ref{fig:contmeas}(c) shows a comparision of the electrical signal for 1~V stimulation and the detected magnetic field $B(t)$  as calculated by deconvolving the optical signal with the magnetometer response [see Methods]. 
The temporal profiles of the electrical signal and the magnetic field look very similar; both show the action potential and a stimulation artifact.
The nerve was stimulated for 50~$\mu$s [inset to Fig.~\ref{fig:pulsedelectricaloptical}] while the magnetic field from the stimulation (Fig.~\ref{fig:contmeas}(c)) lasts for around 0.5~ms. 
This discrepancy is due to the rather slow response time of the magnetometer $T_2 = 0.44$~ms and  low pass filtering of the optical signal in the data analysis.
From the bottom plate of Fig~\ref{fig:contmeas}(c) we conclude that the nerve magnetic field has a 7~pT peak-to-peak amplitude (measured at an average distance of 4.5~mm) and that the nerve conduction velocity is 38(9)~m/s \cite{sup}.
The effective axial ionic current is estimated to be 0.16~$\mu$A  \cite{sup} which is consistent with earlier measurements \cite{Wijesinghe1991}. 

From the data we find the experimental uncertainty $\Delta \abs{B_{\mathrm{exp}}(\Omega)}=5.7$~pT$\cdot$ms for the pulsed mode and a sensitivity of 360~fT/$\rtHz$ in the continuous mode (see Methods section).  
The quantum projection noise limited uncertainty  in the pulsed mode is $\Delta \abs{B_{\mathrm{PN}}(\Omega)}=0.30$~pT$\cdot$ms. In this mode the light is off during the nerve impulse duration $\tau\approx2$~ms which satisfies $\tau \ll \Tdark$.
In the continuous mode, where the $T_2=0.44$~ms matches the nerve impulse bandwidth the PN-limited sensitivity is 29~$\fTrtHz$. 
Thus the projection noise constitutes about $10\%$ of the total experimental uncertainty, the quantum photon shot noise contributes $\approx 50\%$, with the rest due to the uncompensated low frequency classical noise of the probe light and of the atomic spin.

Projection noise dominated sensitivity can be reached by relatively straightforward steps, such as using multiplass vapor cells \cite{Sheng2013prl}, by modest heating  (increasing the temperature from room-temperature $22^\circ$C to the human body temperature $37^\circ$C will increase the sensitivity by a factor of two \cite{SteckCesium}) or by employing a low finesse optical cavity ~\cite{Vasilakis2015}.  Gradiometry with two cells with oppositely oriented spins allows for generation of nonclassical entangled states leading to sensitivity beyond the PN limit  \cite{Wasilewski2010prl} as well as provides additional compensation of the ambient magnetic fields and classical fluctuations of the atomic spins.

In conclusion, we have performed non-invasive detection of nerve impulses from the frog sciatic nerve by measuring the magnetic field generated by the nerve with a room-temperature sensor with near quantum limited sensitivity.
A few mm-sized sensor which is sensitive enough to detect sub-picoTesla fields at a distance of a few millimeters from biological objects makes the magnetometer perfect for medical diagnostics in physiological/clinical areas such as cardiography of fetuses, synaptic responses in the retina, and magnetoencephalography.

%%%%%%%%%%%%%%%%%%%%%%%%%%%%%%%%%%%%%%%
%\cleardoublepage 
\section{Methods}

\subsection{Nerve preparation} 
This study was approved by the Animal Care and Use
Committee of Copenhagen University, and was conducted in accordance with
the Guiding Principles for the Care and Use of Animals in the Field of Physiological
Science. All efforts were made to minimize animal suffering and the number of
animals used.

\begin{figure}[ht]
\centering
\includegraphics[width=0.4\textwidth]{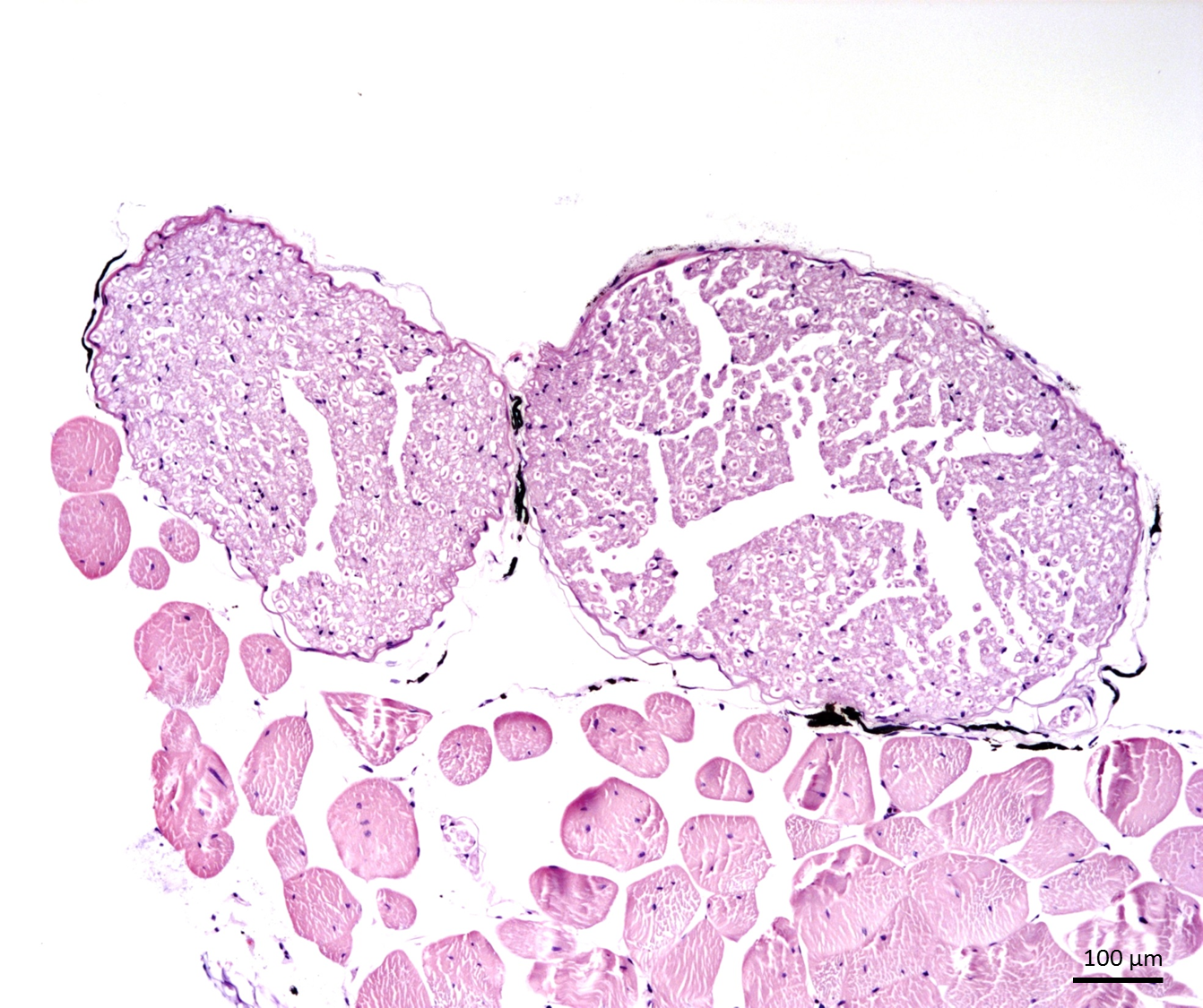}
\caption{Electron micrograph of a frog sciatic nerve.
The micrograph shows the cross section of the frog sciatic nerve on lower femur after its first division. The two nerve bundles are surrounded by a fascia of connective tissue, and they are seen above sections of skeletal muscle. The diameters of the two nerve bundles are 0.59 and 0.44 mm and they each contain 1750 and 950 single axons with a minimal diameter of 7~$\um$. The individual myelinated nerve fibres have an average diameter of 16~$\um$.
}
\label{fig:nerveimage}
\end{figure}

Sciatic nerves were isolated from green frogs (Rana Temporaria). The frogs were
decapitated and the sciatic nerves were isolated from spine and down to the knee. 
The nerves are  7-8~cm long with a diameter of 1.3~mm  in the proximal end and  slightly thinner in the distal end.
In the proximal end there is one bundle that divides twice so distally it is
composed of three bundles.  Figure~\ref{fig:nerveimage} shows an electron micrograph of the nerve where it is seen that the nerve bundles contain a few thousand of axons.

Throughout the dissection and the course of the
experiments, the frog sciatic nerves were kept moist in cold Ringer’s solution (also called saline solution) of 115
mM NaCl, 2.5 mM KCl, 1.8 mM $\mathrm{CaCl_2}$, 1.08 mM $\mathrm{Na_2HPO_4} \cdot \mathrm{2H_2O}$, 0.43 mM
$\mathrm{NaH_2PO_4} \cdot \mathrm{H_2O}$, adjusted to pH 7.1 \cite{Katsuki2006BJP}. Ringer’s solution approximates the ionic
composition of the extracellular fluids of the frog. A high potassium concentration
Ringer’s solution, in which all NaCl was replaced by KCl with final $\mathrm{K^+}$ concentration
being 117mM and $\mathrm{Na^+}$ 0 mM and $\mathrm{Cl^-}$ concentration remaining constant, was used to make nerves inexcitable \cite{Krnjevic1954JPhysiology}.

\subsection{Nerve chamber and electrical recording} 

The nerves are kept in a 3D-printed plastic chamber during the
experiments. The chamber is a 47 x 18 mm block of 8.5 mm height that contains a
longitudinal U-shaped channel with a diameter of 2 mm in which the nerve is placed. 
The channel allows maintenance of a saturated water-vapor atmosphere in order to keep the nerve moist. The front part of the chamber is placed close to magnetometer and it is covered by a microscope glass cover slip of 0.13 mm thickness.

The chamber has 6 circular gold electrodes on each side.  The  electrodes have an outer diameter of 6 mm with the hole inside of 1.5 mm, which fits into the channel. 
On each side, the distance between electrodes is 5 mm.

The nerve was externally stimulated in the proximal end by applying a short 50~$\mu$s
square voltage pulse between two spatially separated electrodes surrounding the
nerve. 
The nerve signal was measured in the distal end as a potential difference between a pair of two electrodes. The electrical signal was amplified 10 times and filtered (using a 3 kHz low pass and a 10 Hz high pass filter) and recorded simultaneously with the magnetic field recordings.

\subsection{Operation of the magnetometer}
The magnetometer is based on optical read-out of spin-polarized atomic vapour. The cesium atoms are prepared by an optical pumping with a pulse of circular polarized light such that the total spin vector $\mathbf{J}=\lr{J_x,J_y,J_z}$ points in the $x$-direction, which is also the direction of a bias field $B_x$ [see Fig.~\ref{fig:setup}(a)]. Any magnetic field perpendicular to $x$-direction (such as the magnetic field from the nerve) will create a transverse spin component $\mathbf{J_\perp}$ which afterwards will precess around the bias magnetic field at the Larmor frequency $\Omega=B_x/\gyromag$ [see Fig.~\ref{fig:setup}(b)], where $\gyromag=2.20 \cdot 10^{10}$~rad/$\lr{\mathrm{s}\cdot \mathrm{T}}$  is the cesium gyromagnetic ratio.  The transverse spin component is detected optically by measuring the polarization rotation of  linearly polarized probe light passing through the vapor cell using a balanced polarimeter.  The magnetic field from a nerve can be detected in two modalities, a pulsed or a continuous one. 
In the continuous mode the pump and probe light is continuously on. In this case, the magnetometer signal $S(t)$ is proportional to the convolution of the magnetic field with the magnetometer response function 
$S(t) \propto \int_{t'=0}^t  \left\{ e^{- \g \lr{t-t'}} \cos \LR{\Omega \lr{t-t'}} \right\} B_y(t')  dt'$, where we assumed the transverse field is along the $y$-direction. 
The relaxation rate $\g =1/T_2$, which is the inverse of the spin-coherence time $T_2$, increases linearly with laser power and is in the limit of low power denoted $\gdark=1/\Tdark$.
In the pulsed mode [see Fig.~\ref{fig:setup}(c)], a pulse of pump light first initializes the atomic spins along the $x$-direction, then the pulse of magnetic field creates a transverse spin component, and finally the spins are detected with a pulse of probe light. In this case, the magnetometer signal is a free induction decay $S(t) =A  \sin \lr{\Omega t + \theta}  e^{-\g t} $, where $\theta$ is a phase and the amplitude $A$ is  proportional to the Fourier component of the magnetic field at the Larmor frequency: $\abs{B\lr{\Omega}}=\abs{\int_{0}^{\tau} B_y(t) e^{-i\Omega t}dt}$, where $\tau$ is the duration of the magnetic field pulse.
For both pulsed and continuous modes, the proportionality constant can be found by applying a known calibration magnetic field.

\begin{figure*}[ht]
\centering
\includegraphics[width=0.75\textwidth]{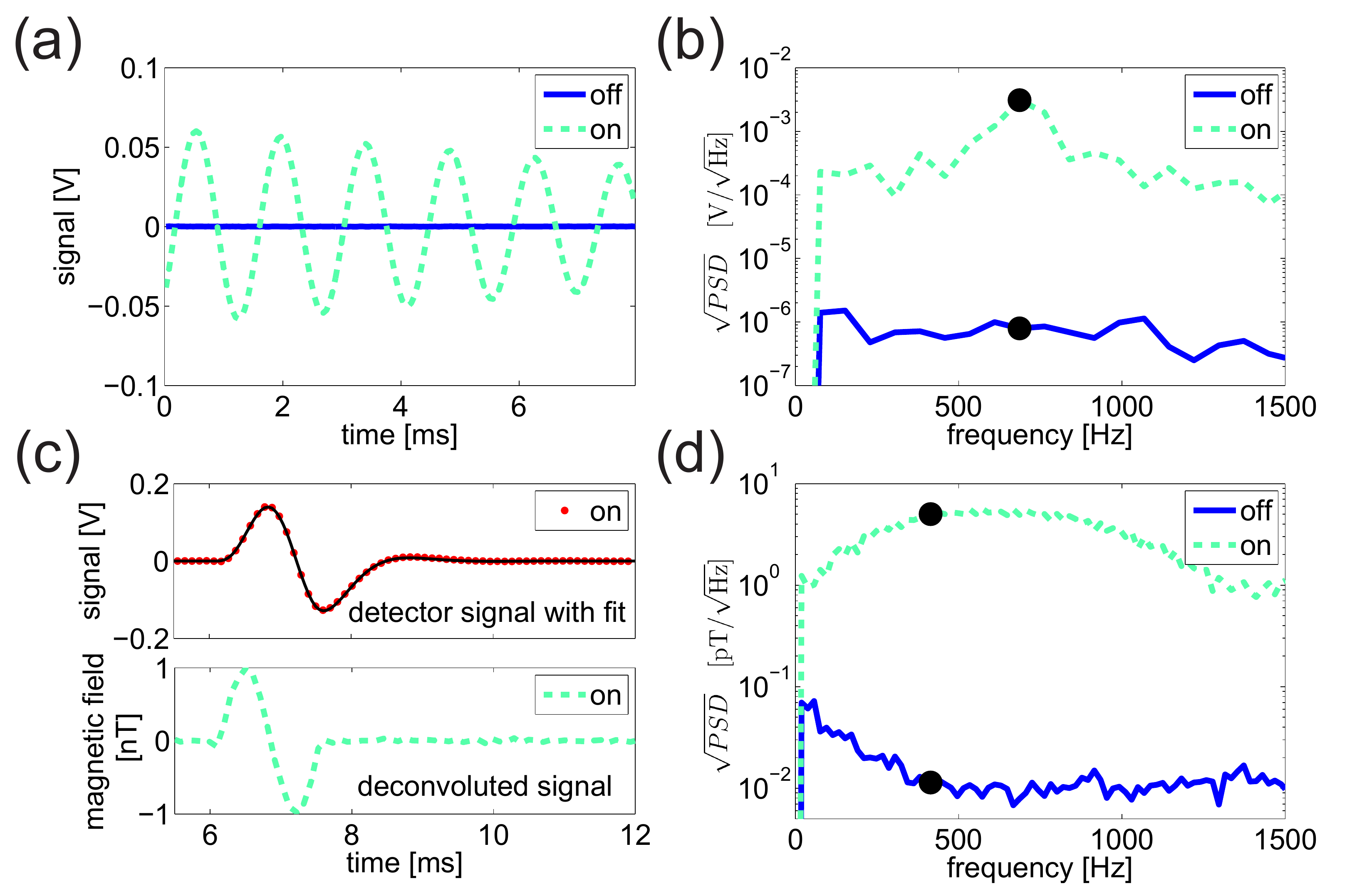}
\caption{Measurements with and without the calibration field. 
(a,b) Pulsed mode: $\Omega=700$~Hz, $\Tdark=15$~ms. The $PSD$ is calculated using the first $8$~ms of the recorded signal.
(c,d) Continuous mode: $\Omega=414$~Hz, $T_2=0.44$~ms. The $PSD$ is calculated using $ 37.1$~ms of the recorded signal.
The Larmor frequencies are marked in (b) and (d) with a dot. }
\label{fig:calib}
\end{figure*}

\subsection{Calibration of the pulsed magnetometer}

The magnetometer is calibrated by applying a known magnetic field.
This calibration field is produced by a coil positioned inside the magnetic shield; 
the field points in the $z$-direction and has the temporal shape of a single sinusoidal oscillation 
$B_{\mathrm{cal}}(t)= B_{\mathrm{cal}}  \sin \lr{\Omega_{\mathrm{cal} } t}$
with  amplitude $B_{\mathrm{cal}} =1$~$ \mathrm{nT}$,  frequency $\Omega_{\mathrm{cal}}  = 2\pi \cdot 700$~Hz
and Fourier component $\abs{B_{\mathrm{cal}}(\Omega_{\mathrm{cal}})}= \pi B_{\mathrm{cal}} /\Omega_{\mathrm{cal}}=0.71~$nT$\cdot$ms.

In the pulsed mode, the calibration field is applied in between the pump and probe pulses.
The recorded magnetometer signal (the free induction decay)  is shown in Figure~\ref{fig:calib}(a). 
The  spectrum (calculated as the square-root of the power spectral density $\sqrt{PSD}$) is peaked at the Larmor frequency of the atoms $\Omega=2\pi\cdot700$~Hz [Fig.~\ref{fig:calib}(b)].
The peak amplitude  is proportional to the magnetic field Fourier component: 
$\sqrt{PSD\lr{\Omega}} \propto \abs{B\lr{\Omega}}$.
The proportionality constant can be calculated from the data in Fig.~\ref{fig:calib}(b) and the known Fourier component of the calibration field. 
With this calibration we can calculate the Fourier component 
$|B_\mathrm{nerve}(\Omega)|$ of the nerve magnetic field [see Fig.~\ref{fig:pulsedelectricaloptical}(f)] from the measured peak values $\sqrt{PSD(\Omega)}$ [see Fig.~\ref{fig:pulsedelectricaloptical}(e)].

Figures~\ref{fig:calib}(a) and \ref{fig:calib}(b) also show the magnetometer signal without the applied calibration field. The signal was averaged 1000 times before recorded. The noise at the Larmor frequency is 3900 times smaller than the signal obtained with the calibration field, i.e., the calibration field is detected with a  $SNR=3900$ corresponding to a minimal detectable magnetic field Fourier component 
$0.71$~$\mathrm{nT}\cdot \mathrm{ms}/3900=0.18$~$\mathrm{pT}\cdot\mathrm{ms}$ using 1000 averages. The single shot  minimal detectable Fourier component is then $5.7$~$\mathrm{pT}\cdot\mathrm{ms}$.

\subsection{Calibration of the continuous magnetometer}
In the continuous mode, the lasers are on during the applied calibration field. In this case the magnetometer signal is proportional to the convolution of the magnetic field with the magnetometer response function.
Figure~\ref{fig:calib}(c) top  shows the detected signal together with a fit to the function
$S(t) \propto \int_{t'=0}^t  \left\{ e^{- \g\lr{t-t'}} \sin \LR{\Omega \lr{t-t'}} \right\} B_z(t')  dt'$ from which we can determine the Larmor frequency and the coherence time.
Using the fitted parameters we can perform numerical deconvolution and by scaling with the amplitude of the calibration field we can obtain the magnetic field as a function of time $B_z(t)$ [Figure~\ref{fig:calib}(c) bottom ]. We see that the deconvolution procedure works well as the deconvolved signal resembles a single sinusoidal oscillation.

Figure~\ref{fig:calib}(d) shows the calculated noise spectrum of the deconvoluted signal. It also shows the noise spectrum when the calibration field was off. In this case the noise at the Larmor frequency is 11.4~fT/$\rtHz$ when 1000 averages is used. The single shot magnetic field sensitivity is then 360~fT/$\rtHz$.

\section*{Acknowledgements}
We would like to thank J\"{o}rg Helge M\"{u}ller for helpful discussions, Bing Chen for help during the initial stages of the experiments, and Hannelouise Kissow for taking the electron micrograph image of the nerve. This work was supported by the ERC grant INTERFACE and by the DARPA project QUASAR. K. J. was supported by the Carlsberg Foundation. R. A. T.  was supported by CNPq/Brazil.
%%%%%%%%%%%%%%%%%%%%%%%%%%%%%%%%%%%%%%%%%%%%%%%%%%

%%%%%%%%%%%%%%%%%%%%%%%%%%%%%%%%%%%%%%%%%%%%%%%%%%
\cleardoublepage 

%% for citing papers
%\renewcommand*{\citenumfont}[1]{S#1}
%\renewcommand*{\bibnumfmt}[1]{[S#1]}
%% for labeling figures and equations
%\renewcommand{\thefigure}{S\arabic{figure}} 
%\renewcommand{\theequation}{S\arabic{equation}} 

\section{\underline{SUPPLEMENTARY INFORMATION}}

\section{Magnetometer principle}
The total spin of the atomic ensemble is defined as $\mathbf{J}=\lr{J_x,J_y,J_z}$. Here $\vecJ$ is a quantum operator, and the components have the commutation relation $\left[J_y,J_z \right] = i J_x$. $\vecJ$ is here defined as being unitless and equals the total angular momentum divided by the reduced Planck constant $\hbar$.
The equations of motion for the spin vector can be derived using the Heisenberg equation of motion  
\begin{equation}
\dot{\vecJ}(t) = \frac{1}{i \hbar}  \left[ \vecJ(t) ,\HB \right].
\end{equation}
The dot denotes the time-derivative and the  bracket denotes the commutator. 
The Hamiltonian describing the coupling  between the spin and the magnetic field is
\begin{equation}
\HB = \hbar \gyromag \vecB \cdot \vecJ,
\end{equation}
where  $\gyromag$ is the gyromagnetic ratio which for the cesium atom in the $F=4$ ground state equals 
$2.2 \times \nobreak 10^{10}$~rad/(sec$\cdot$Tesla).
In vector form the equation of motion reads
\begin{equation}
\dot{\vecJ}(t) = \gyromag \vecJ \times \vecB .
\label{eq:Jdot}
\end{equation}

In the experiment, the atoms are spin-polarized in the $x$-direction and located in a static magnetic field $B_x$  pointing in the $x$-direction.
In the presence of a small time-dependent magnetic field $B_y(t)$ or $B_z(t)$ pointing in  the  $y$- or $z$-direction, the spin vector will acquire a transverse component 
$\vecJ_\perp = \left( J_y,J_z\right) = \abs{\vecJ_\perp} \lr{\cos \theta, \sin \theta}$.
We will assume that $J_x$ is large compared to $J_y$ and $J_z$, and that $J_x$ is independent of time. 
We now  introduce spin operators $J'_y$ and $J'_z$ rotating at the Larmor frequency $\Omega =\gyromag B_x$:
\begin{equation}
\begin{pmatrix}
J'_y\\
J'_z
\end{pmatrix} 
=
\begin{pmatrix}
 \cos \Omega t & \sin \Omega t\\
-\sin \Omega t  & \cos \Omega t
\end{pmatrix}
\begin{pmatrix}
J_y\\
J_z
\end{pmatrix} .
\end{equation}
In the rotating frame, the equations of motion read
\begin{IEEEeqnarray}{lCl} 
\dot{J'_y}(t) &=& \gyromag J_x  \left[\cos\lr{\Omega t}B_z(t) - \sin\lr{\Omega t}B_y(t)  \right] \nonumber \\
                     &  &  - \g J'_y(t)  + \sqrt{2\g}  F_y(t)   , \\
\dot{J'_z}(t) &=& -\gyromag J_x  \left[\sin\lr{\Omega t}B_z(t) + \cos\lr{\Omega t}B_y(t)  \right] \nonumber \\
                     &  &   - \g J'_z(t)  + \sqrt{2\g} F_z(t)   .
\end{IEEEeqnarray}
The transverse spin component will eventually decay, and we have therefore added decay terms in the above equations.
The decay rate is denoted by $\g$ and the associated decay time is $T_2=1/\g$. 
We also added Langevin noise operators $F_y(t)$ and $F_z(t)$ with zero mean values and correlation functions $\braket{F_y(t) F_y(t')} = \var \lr{F_y} \delta(t-t')$, $\braket{F_z(t) F_z(t')} = \var \lr{F_z} \delta(t-t')$ and $\braket{F_y(t) F_z(t')} = 0$, where $\var(F_y)= \var(F_z)= \abs{J_x}/2$ and $\delta(t-t')$ is the Dirac delta-function.
These equations can be integrated and the solutions are
\begin{IEEEeqnarray}{lll} % can use lCr
J'_y(t)  =   e^{-\g t} J'_y(0) + \sqrt{2\g} \int_{t'=0}^t e^{-\g \lr{t-t'}} F_y(t') dt'&&  \nonumber\\
 +\gyromag  J_x \int_{t'=0}^t  e^{-\g \lr{t-t'}}
\left[ \cos\lr{\Omega t'} B_z(t') -\sin\lr{\Omega t'} B_y(t')    \right] dt' , && \nonumber \\
&&\\
J'_z(t)  =   e^{-\g t} J'_z(0) + \sqrt{2\g} \int_{t'=0}^t e^{-\g \lr{t-t'}} F_z(t') dt && \nonumber  \\
-\gyromag  J_x \int_{t'=0}^t  e^{-\g \lr{t-t'}}
\left[ \sin\lr{\Omega t'} B_z(t') +\cos\lr{\Omega t'} B_y(t')    \right] dt' . && \nonumber \\
&&
\end{IEEEeqnarray}
From these equations we can calculate  the mean values and noise proporties of the transverse spin components as a function of time.

\subsection{Free precession}
Assume that the transverse spin has some mean value at $t=0$ and that it is then left free to precess. At a later time $t$ the transverse spin component in the rotating frame is
\begin{equation}
 \braket{\vecJ'_\perp(t)} =  \braket{\vecJ_\perp(0)} e^{-\g t} .
\label{eq:Jzperpdecay}
\end{equation}
We see that the mean value decays in time. In the lab frame the transverse spin will perform a damped oscillation.

\subsection{Atomic response to a pulse of magnetic field}
Consider the case where a magnetic field $B_z(t)$ is applied for a duration $\tau$. We assume that $\tau \ll T_2$ such that any decay of the spin components can be neglegted. If initially the transverse spin component is zero
$\bracket{\vecJ_{\perp}(0)}=0$, we find
\begin{equation}
\bracket{\vecJ'_{\perp}(\tau)} = \gyromag J_x \left(  \Re [B_z(\Omega)], \Im [B_z(\Omega)] \right)
\end{equation}
and 
\begin{equation}
\abs{\braket{\vecJ'_\perp(\tau)} }= J_x \gyromag \abs{B_z(\Omega)}.
\label{eq:JperpBz}
\end{equation} 
Here we have defined  the Fourier component of the magnetic field at the Larmor frequency as
\begin{equation}
B(\Omega) = \int_{t'=0}^\tau B(t')  e^{-i\Omega t'} dt.
\end{equation}
Similarly, if the magnetic field is applied along the $y$-direction instead, the transverse spin component will be 
\begin{equation}
\abs{\braket{\vecJ'_\perp(\tau)} }= \gyromag J_x  \abs{B_y(\Omega)}.
\label{eq:JperpBomega}
\end{equation}
We see that magnetic fields in $y$- and $z$-directions have similar effects on the spins: the fields create transverse spin components with lengths proportional to the Fourier components of the magnetic fields at the Larmor frequency.

For the specific case of a sinusoidal magnetic field $B_z(t)=B_0 \sin\lr{\Omega t}$ applied for one period of oscillation $\tau=2\pi/\Omega$, we find $\abs{B_z(\Omega)}=\pi B_0/\Omega$ and
$\abs{\braket{\vecJ'_\perp(\tau)} }= \nobreak \gyromag J_x  \lr{\pi B_0/\Omega}$.

\subsection{Projection noise limited detection}
The measurement of the transverse spin component is fundamentally limited by the spin-projection noise originating from the Heisenberg uncertainty principle. This uncertainty is $\Delta \abs{ J_\perp}=\sqrt{J_x/2}$. By equating the created mean value given by Eq.~(\ref{eq:JperpBomega}) to the projection noise we find the uncertainty on the magnetic field Fourier component due to the projection noise:
\begin{equation}
\Delta \abs{\BPN(\Omega)} = 1/\lr{\gamma \sqrt{2J_x}}.
\label{eq:BPN}
\end{equation}
We can also calculate the uncertainty on the amplitude of an oscillating magnetic field due to the projection noise.
For a sinusoidal magnetic field with total duration $\tau$ equal to an integral multiple of the Larmor period, the amplitude $B_0$ is related to the Fourier component by $\abs{B(\Omega)}= B_0 \tau/2$.
From this and Eq.~(\ref{eq:BPN}) we find the projection noise limited uncertainty on the amplitude:
\begin{equation}
\Delta \BPN = 1/\lr{\gamma \sqrt{J_x/2} \tau},
\end{equation}
which is often called the minimal detectable field.
The magnetic field sensitivity can be found by multiplying $\Delta \BPN$ with the square-root of the total measurement time $\sqrt{\Ttot}$ and setting $\Ttot=\tau=T_2$:
\begin{equation}
\Delta \BPN \sqrt{\Ttot} \sim 1/ \lr{\gyromag \sqrt{T_2 J_x /2}}.
\end{equation}

\subsection{Measuring the atomic signal}
The atomic spin can be measured optically. Assume that a linearly polarized pulse of light is propagating in the $z$-direction  through the atomic ensemble. The polarization of the light will be rotated by an angle proportional to $J_z$ due to the Faraday paramagnetic effect. 
The polarization of the light is described using Stokes operators $S_x(t)$, $S_y(t)$, and $S_z(t)$ which have the unit of 1/time.
Here $S_x(t) = \left[\Phi_x(t)-\Phi_y(t)\right]/2$ equals one half the difference in photon flux of $x$- and $y$-polarized light. $S_y(t)$ refer to the differences of $+45^\circ$ and $-45^\circ$ polarized light, and $S_z(t)$ to the differences of right hand and left hand circular polarized light.
Assuming that the input light before the atomic ensemble is either $x$ or $y$-polarized (such that $S_x(t)$ is a large quantity) and that the rotation angle is small, the output light after the atomic ensemble can be described by the equation
\begin{eqnarray}
\Syout (t) & = &  \Syin (t) +a S_x(t) J_z(t) \nonumber \\
                & =  &  \Syin (t) +a S_x(t) \left[ \sin \lr{\Omega t}J'_y(t) +  \cos \lr{\Omega t}J'_z(t) \right] . \nonumber \\
\end{eqnarray}
The parameter $a$ describes the coupling strength between the atoms and the light \cite{Jensen2011phd}.
The Stokes operator  $\Syout(t)$ can be measured with polarization homodyning.
There are several ways that one can extract information about the transverse spin components and therefore about the magnetic field from the measured signal. One can for instance measure the mean value $\braket{\Syout}$ or the power spectral density of the signal.
The power spectral density ($PSD$) for a function $x(t)$ is defined as
\begin{eqnarray}
S_{xx}(\omega) & = & \frac{1}{T}\bracket{\abs{\int_{t=0}^T x(t) e^{-i \omega t} dt}^2} \nonumber \\
&& \frac{1}{T} \int_{t=0}^T \int_{t=0}^T  \bracket{x(t)x(t')} e^{-i \omega \lr{t-t'}} dt dt' \nonumber \\
\end{eqnarray}
We will show below that the $PSD$ of $\braket{\Syout(t)}$ is proportional to the amplitude squared of the applied magnetic field.

\subsection{Detection of a pulse of magnetic field}
Assume that a pulse of magnetic field $B_z(t)$ of duration $\tau$ is applied from $t=-\tau$ to $t=0$. After the pulse, the spins have acquired a non-zero transverse spin component $\braket{\vecJ'_\perp(0)} \propto \abs{B_z(\Omega)}$ as given by Eq.~(\ref{eq:JperpBz}). 
At $t=0$ the spin will continue to precess until it decays as described by Eq.~(\ref{eq:Jzperpdecay}).
This spin vector can be measured using a pulse of light with duration $T$ and starting at the time when the magnetic field pulse ends.
The mean value of the measured signal is
\begin{IEEEeqnarray}{lcl} % can use lCr
\braket{\Syout (t)} &=& 
a S_x(t) \left[ \sin \lr{\Omega t} \braket{J'_y(0)} +  \cos \lr{\Omega t} \braket{J'_z(0)} \right] e^{-\g t}  \nonumber \\
& = &   a S_x(t) \abs{\braket{\vecJ_\perp'(0)}}  \sin \lr{\Omega t + \theta}  e^{-\g t} , 
\label{eq:SyoutJperp}
\end{IEEEeqnarray}
where $\theta$ is the polar angle of $\braket{\vecJ'_\perp(0)}$.
The amplitude of the transverse spin vector  can  be extracted from the measurement by,  for instance, a fit of the experimental data to Eq.~(\ref{eq:SyoutJperp}).
Alternatively, one can calculate the $PSD$ of the signal.
For $x(t)=A \sin \lr{\Omega t + \theta}  e^{-\g t}$ we calculate that the peak value of the $PSD$ is
\begin{equation}
S_{xx}(\Omega) = \abs{A}^2 \left[ \frac{\lr{1-e^{-\g T}}^2}{4\g^2} + \epsilon(\theta,\g,\Omega,T) \right],
\end{equation}
where the second term $\epsilon(\theta,\g,\Omega,T)$ is much smaller than the first term for our experimental parameters.
We see that
\begin{equation}
S_{xx}(\Omega) \propto \abs{\braket{\vecJ'_\perp(0)}}^2 \propto \abs{B_z(\Omega)}^2, 
\end{equation}
and  that the Fourier component of the magnetic field at the Larmor frequency can be extracted from the peak value of the $PSD$.

\subsection{Continuous recording of the magnetic field}
We will now discuss how one can measure the magnetic field as a function of time.
Assume that a magnetic field $B_z(t)$ is applied and that light is continuously monitoring the atomic spin.
If $\braket{\vecJ_\perp(0)}=0$, then at a later time
\begin{equation}
\braket{J_z(t)} =  \gyromag J_x \int_{t'=0}^t 
e^{-\g \lr{t-t'}}
\sin \LR{\Omega \lr{t-t'}}B_z(t')  dt' .
\end{equation}
The mean value of the measured signal will be
\begin{eqnarray}
\braket{\Syout (t)}  =  a S_x(t) \braket{J_z(t)} .
\end{eqnarray}
From this we see that the measured signal $\braket{\Syout (t)} $ is proportional to the convolution of the magnetic field $B_z(t)$ with the function $\LR{\sin \lr{\Omega t} e^{-\g t}}$. 
Similarly, if the transverse magnetic field is pointing in the $y$-direction, the signal is proportional to the convolution of $B_y(t)$ with the function 
$\LR{- \cos \lr{\Omega t} e^{-\g t}}$. The magnetic field as a function of time can be extracted from the measured data using numerical deconvolution.

%%%%%%%%%%%%%%%%%%%%%%%%%%%%%%%%%%%%%%%%%%%%%%%%%%%%%%%

\begin{figure*}[ht]
\centering
\includegraphics[width=1\textwidth]{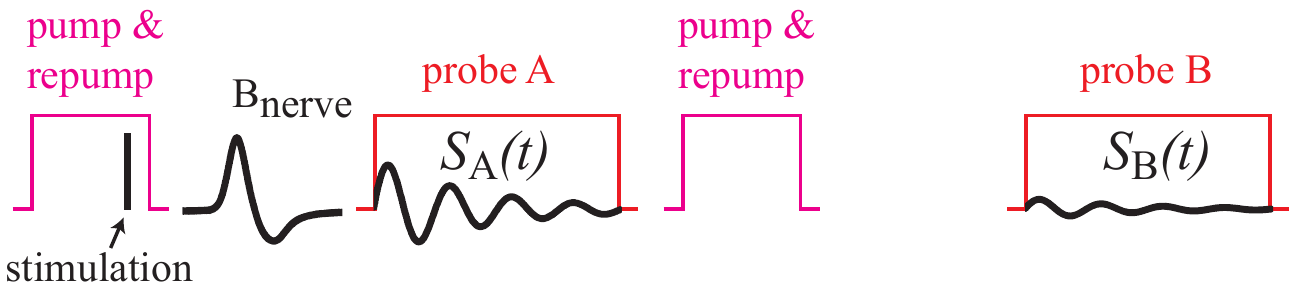}
\caption{Pulse sequence for measuring the nerve impulse.}
\label{fig:pulseseq}
\end{figure*}

\section{Experimental procedure}
Three lasers denoted pump, repump and probe are used in the experiment. The pump laser is on resonance with the cesium $F=4 \rightarrow F'=4$ D1 transition and has the wavelength 895~nm. The repump laser is on resonance with the cesium $F=3 \rightarrow F'=2,3,4$ D2 transitions (all are within the Doppler linewidth) and has the wavelength 852~nm. These two lasers are used for optical pumping of the cesium atoms into the $F=4,m=4$ hyperfine sublevel and are thereby creating a high spin-polarization of the cesium vapor.
The probe laser is 1.6~GHz higher in frequency than the cesium $F=4 \rightarrow F'=5$ D2 transition and has the wavelength 852~nm. The 1.6~GHz detuning is much larger than both the natural linewidth (5~MHz FWHM) and the Doppler linewidth (380~MHz FWHM) such that negligible absorption occurs.

The pulse sequence in Fig~\ref{fig:pulseseq} is used for detection of the magnetic field from the nerve impulse $B_{\mathrm{nerve}}$. 
The atoms are first optically pumped using pump and repump light, then the magnetic field is present, and finally the atoms are measured using probe pulse A. 
The optically detected signal $S_A(t)$ will be a free induction decay as seen in Fig~\ref{fig:pulseseq}. Due to misalignment of the pump and repump laser beams with respect to the bias field $B_x$ (see Fig.~\ref{fig:setup} in the main text), one may observe a free induction decay $S_B(t)$ [see probe pulse B in Fig.~\ref{fig:pulseseq}] even if there is no magnetic field. The pump/repump and probe pulses are therefore repeated  and the signals from probe pulses A and B are subtracted giving the magnetometer signal $S(t)= S_A(t)-S_B(t)$. The amplitude of this magnetometer signal will be proportional to the Fourier component of the magnetic field at the Larmor frequency $\abs{B(\Omega)}$.

\section{Conduction velocity}
The nerve conduction velocity can be calculated by dividing the distance from the stimulation electrodes to the recording site [5(1)~cm for optical recording and 7(1)~cm for electrical recording] by the time interval  between the stimulus artifact and the peak of the nerve impulse. The earlier arrival of the nerve impulse for optical recording compared to electrical recording [1.3(2)~ms compared to 1.9(1)~ms]  is consistent with the magnetometer being positioned in between the stimulating and recording electrodes. 
From the measurements Fig.~\ref{fig:contmeas}(c) we calculate the conduction velocity of 38(9)~m/s and 37(6)~m/s for optical and electrical recording.

\section{Estimate of the axial ionic current}
The detected magnetic field is created by axial ionic currents inside the nerve bundle. There is a forward current inside the axons and a return current  outside the axons. The magnetic fields from the forward and return currents can cancel each other, the exact degree of cancellation depends on the anatomy of the nerve, and the geometry of the experiment, such as the size of the magnetic field sensor and the distance from the nerve to the sensor. 

We can estimate the axial current in the nerve from our magnetic field measurements.
We use a simple model, where we assume that the ionic current is concentrated at the center of the nerve, and that the nerve produces a magnetic field similar to that of an infinitely long conducting wire. In the experiment, we detect the magnetic field averaged over the volume of the spherical vapor cell. We calculate that the average magnetic field is equal to the field at the center of the vapor cell. 
The magnetic field from an infinitely long wire is
$\abs{B}=\mu_0 I/\lr{2 \pi r}$, where $\mu_0$ is the magnetic permeability, $I$ is the current, and $r$ is the radial distance from the wire.
Using $r=4.5$~mm for the distance from the center of the nerve to the center of the vapor cell, we calculate that a current of 0.16~$\mu$A will produce a magnetic field of 7~pT.

Our estimate of 0.16~$\mu$A is smaller than the 0.4~$\mu$A which was estimated in previous work  on the frog sciatic nerve \cite{Wijesinghe1991}. This is expected as in that work,  the nerve was put in a large container with saline solution and the magnetic field was measured by a coil with the nerve inside it, such that a large part of the return current could flow without being detected.

%\bibliography{BIBQuantop2} 

\begin{thebibliography}{10}

\bibitem{Wikswo1980}
J.~P. Wikswo, J.~P. Barach, and J.~A. Freeman.
\newblock Magnetic field of a nerve impulse: first measurements.
\newblock {\em Science}, 208(4439):53--55, 1980.

\bibitem{Budker07}
Dmitry Budker and Michael Romalis.
\newblock Optical magnetometry.
\newblock {\em Nature Physics}, 3:227, 2007.

\bibitem{Kominis2003nature}
I.~K. Kominis, T.~W. Kornack, J.~C. Allred, and M.~V. Romalis.
\newblock A subfemtotesla multichannel atomic magnetometer.
\newblock {\em Nature}, 422:596, 2003.

\bibitem{Alem2015PhysMedBio}
O.~Alem, T.~H. Sander, R.~Mhaskar, J.~LeBlanc, H.~Eswaran, U.~Steinhoff,
  Y.~Okada, J.~Kitching, L.~Trahms, and S.~Knappe.
\newblock Fetal magnetocardiography measurements with an array of
  microfabricated optically pumped magnetometers.
\newblock {\em Physics in Medicine and Biology}, 60(12):4797, 2015.

\bibitem{Sander2012biomedexpress}
T.~H. Sander, J.~Preusser, R.~Mhaskar, J.~Kitching, L.~Trahms, and S.~Knappe.
\newblock Magnetoencephalography with a chip-scale atomic magnetometer.
\newblock {\em Biomed. Opt. Express}, 3(5):981--990, 2012.

\bibitem{Alem2014neuro}
O.~Alem, A.~M. Benison, D.~S. Barth, J.~Kitching, and S.~Knappe.
\newblock Magnetoencephalography of epilepsy with a microfabricated atomic
  magnetrode.
\newblock {\em The Journal of Neuroscience}, 34(43):14324, 2014.

\bibitem{Wasilewski2010prl}
W.~Wasilewski, K.~Jensen, H.~Krauter, J.~J. Renema, M.~V. Balabas, and E.~S.
  Polzik.
\newblock Quantum noise limited and entanglement-assisted magnetometry.
\newblock {\em Phys. Rev. Lett.}, 104:133601, 2010.

\bibitem{Bison2009apl}
G.~Bison, N.~Castagna, A.~Hofer, P.~Knowles, J.-L. Schenker, M.~Kasprzak,
  H.~Saudan, and A.~Weis.
\newblock A room temperature 19-channel magnetic field mapping device for
  cardiac signals.
\newblock {\em Applied Physics Letters}, 95:173701, 2009.

\bibitem{sup}
See Supplementary Material.

\bibitem{Balabas2010oe}
M.~V. Balabas, K.~Jensen, W.~Wasilewski, H.~Krauter, L.~S. Madsen, J.~H.
  M\"{u}ller, T.~Fernholz, and E.~S. Polzik.
\newblock High quality anti-relaxation coating material for alkali atom vapor
  cells.
\newblock {\em Opt. Express}, 18(6):5825, 2010.

\bibitem{Balabas2010prl}
M.~V. Balabas, T.~Karaulanov, M.~P. Ledbetter, and D.~Budker.
\newblock Polarized alkali-metal vapor with minute-long transverse
  spin-relaxation time.
\newblock {\em Phys. Rev. Lett.}, 105:070801, 2010.

\bibitem{Krnjevic1954JPhysiology}
K.~Krnjevic.
\newblock Some observations on perfused frog sciatic nerves.
\newblock {\em The Journal of Physiology}, 123(2):338--356, 1954.

\bibitem{Wijesinghe1991}
R.~S. Wijesinghe, F.~L.~H. Gielen, and J.~P. Wikswo.
\newblock A model for compound action potentials and currents in a nerve bundle
  {I}: The forward calculation.
\newblock {\em Annals of Biomedical Engineering}, 19(1):43, 1991.

\bibitem{Sheng2013prl}
D.~Sheng, S.~Li, N.~Dural, and M.~V. Romalis.
\newblock Subfemtotesla scalar atomic magnetometry using multipass cells.
\newblock {\em Phys. Rev. Lett.}, 110:160802, 2013.

\bibitem{SteckCesium}
Daniel A. Steck, “Cesium D Line Data,” available online at
  http://steck.us/alkalidata (revision 2.1.4, 23 December 2010).

\bibitem{Vasilakis2015}
G.~Vasilakis, H.~Shen, K.~Jensen, M.~Balabas, D.~Salart, B.~Chen, and E.~S.
  Polzik.
\newblock Generation of a squeezed state of an oscillator by stroboscopic
  back-action-evading measurement.
\newblock {\em Nature Physics}, 11:389, 2015.

\bibitem{Katsuki2006BJP}
R.~Katsuki, T.~Fujita, A.~Koga, T.~Liu, T.~Nakatsuka, M.~Nakashima, and
  E.~Kumamoto.
\newblock Tramadol, but not its major metabolite (mono-\textit½{O}-demethyl
  tramadol) depresses compound action potentials in frog sciatic nerves.
\newblock {\em British Journal of Pharmacology}, 149(3):319--327, 2006.

\bibitem{Jensen2011phd}
K.~Jensen.
\newblock {\em Quantum information, entanglement and magnetometry with
  macroscopic gas samples and non-classical light}.
\newblock PhD thesis, University of Copenhagen, 2011.

\end{thebibliography}
%\bibliographystyle{unsrt} 

\end{document}